\documentclass[preprint,aps,amsmath,amssymb,nofootinbib,12pt]{revtex4}
\usepackage{epsfig}
\usepackage{graphicx,subfigure}
\usepackage{slashed}
\usepackage{float}

\usepackage{multirow,dcolumn,color}
\usepackage{amsmath}

\usepackage{amssymb}
\usepackage{enumerate}
\usepackage{verbatim}
\usepackage[sc]{titlesec}

\graphicspath{{fig/}}
\newcommand{\missE}{\slashed E}
\def\beq{\begin{equation}}
\def\eeq{\end{equation}}
\def\bea{\begin{eqnarray}}
\def\eea{\end{eqnarray}}

\newcommand{\WWd}{W_{\mu\nu}}

\begin{document}
\title{
Prospects for detecting axion-like particles via the decay $Z\rightarrow af\bar{f}$ at future $Z$ factories}
\author{Chong-Xing Yue}
\thanks{cxyue@lnnu.edu.cn}
\author{Shuo Yang}
\thanks{shuoyang@lnnu.edu.cn(corresponding author)}
\author{Han Wang}
\thanks{wangwanghan1106@163.com}
\author{Nan Zhang}
\thanks{zn1517499457@163.com}
\affiliation{Department of Physics, Liaoning Normal University, Dalian 116029, China}

\begin{abstract}
We investigate the prospects of detecting axion-like particles (ALPs, dubbed as "a") via the decay $Z\rightarrow a f\bar{f}$ at future $Z$ factories. Considering the decay channels $a\rightarrow \mu^+ \mu^-$ and $a\rightarrow b \bar{b}$ , four types of signals $\mu^+ \mu^- \slashed E$, $b b\slashed E$, $e^+ e^- \mu^+ \mu^-$  and $e^+ e^- b b$  are explored. We demonstrate that these channels are promising for detecting ALPs at $Z$ factories and obtain the sensitivity bounds on the couplings $g_{aZZ}$ and $g_{a\gamma Z}$.
\end{abstract}

\maketitle

\section{Introduction}
Many new physics scenarios beyond the standard model (SM) predict the existence of axion-like particles (ALPs), which are generalizations of QCD axions proposed as a solution to the strong CP problem~\cite{CP} and are neutral pseudoscalar particles of the broken global symmetry at high scale~\cite{HS}. Their masses are natural small compared to the broken scale $\Lambda$. Unlike the QCD axion, the mass and the couplings of ALP might be independent free parameters to be probed experimentally. This property makes ALPs have a much wider parameter space and hence generate rich phenomenology at low-energy and high-energy experiments~\cite{EFTandCollider,ALPCollider,ALPMevto90,ALPCollider2,ALPee,ALPbosons,ALPtriboson,ALPFCNC,ALPEDM,Flavorcons,ALPFlavor,ALPLFV,ALPLFV2,ALPBelle,agg,nonresggF,nonresVBS,DMwithZ,Z2axgamma,
Zhang1,Guo,avv,Wu,Zhang2,ALPLHCb}. Thus, there are many experimental prospects to search for ALPs, for recent reviews see:~\cite{ALPEXP}, which is the primary coverage of searching for new physics.

As ALPs coming from the breaking of a global symmetry at high energy scale, their interactions with ordinary particles can be suitably described via an effective Lagrangian and studied in effective field theory (EFT) framework~\cite{GEO,EFTandCollider,ALPCollider}. The constraints on the effective couplings of ALPs to the SM fermions or bosons have been widely studied~\cite{EFTandCollider,ALPCollider,ALPMevto90,ALPCollider2,ALPee,ALPbosons,ALPtriboson,ALPFCNC,ALPEDM,Flavorcons,ALPFlavor,ALPLFV,ALPLFV2,ALPBelle,agg,nonresggF,nonresVBS} using various experimental data from particle physics, astroparticle physics and cosmology. The severity of the constraints depends on the ALP mass range considered. Generally speaking, the most stringent limits on ALP couplings arise from cosmological and astrophysical bounds for very light ALPs with mass below the MeV~\cite{EFTandCollider,ALPCollider}. For ALPs with MeV to hundreds of GeV scale masses, collider experiments become relevant and the constraints are alleviated. If the light ALP is long-lived, it would leave a missing energy signature in the detector and the mono-$X$ ($X=\gamma,~Z,~W^{\pm}$) searches at colliders are targeted for it~\cite{EFTandCollider}. ALPs with larger masses or larger couplings become short-lived and can decay inside the detector. For example, the electroweak gauge boson $Z$ decaying into two or three photons at $e^{+}e^{-}$ colliders and hadron colliders can generate constrains on the coupling of ALP to photons~\cite{ALPMevto90}. The current limits provide valuable information for direct searching ALPs in running or future high- and low-energy experiments.

It is well known that the electroweak gauge boson $Z$ can be copiously  produced at high-energy collider experiments. Two proposed $Z$ factories, the Circular Electron-Positron collider (CEPC)~\cite{CEPCPCDR,CEPCCDR} and the Future Circular Collider (FCC-ee)~\cite{FCCee1,FCCeeCDR,FCCCDR}, are assumed to produce up to $10^{12}$ $Z$ bosons. It is possible for the high-precision measurements and the observations of rare processes by these large data samples, which will open opportunities for ALPs detecting. Refs.~\cite{DMwithZ,Z2axgamma,Chang} find that the future $Z$ factories will play an important role in uncovering some extensions of the SM. Previous studies on searching for ALPs in exotic decay of $Z$ are mainly focused on $Z\rightarrow a\gamma$ channel and the following decay $a\rightarrow \gamma \gamma $~\cite{EFTandCollider,ALPCollider,ALPMevto90,Z2axgamma}.
In this paper, the possibilities of detecting ALPs via the exotic decay $Z \rightarrow a f\bar{f}$ followed by $a$ decaying into fermions are investigated. The mass range of ALPs from 5 GeV to 70 GeV is considered.

The structure of this paper is as follows. After summarizing the effective description of ALP interactions and experimental constraints in section II, we perform a detailed analysis on the probability of detecting ALP via the process $Z \rightarrow af\bar{f}$ with $a$ decaying into $\mu^+ \mu^-$ and $b\bar{b}$ at future $Z$ factories. Finally, summaries and discussions are given in section IV.

\section{Effective interactions of ALP}
The ALP $a$ is regarded as CP-odd boson, is a singlet under the SM gauge group. The effective interactions of ALP with the SM particles can be described by the effective Lagrangian. The most general effective Lagrangian including operators of dimension up to 5 in electroweak sector is given by~\cite{GEO,EFTandCollider,ALPCollider}

\bea
\label{lagrangian}
\mathcal{L}_{\text {eff }}^{D \leq 5} &=& \frac{1}{2}\left(\partial_{\mu} a\right)\left(\partial^{\mu} a\right)-\frac{m_{a}^{2}a^{2}}{2} +\frac{\partial^{\mu} a}{f_a} \sum_{\substack{\psi=Q_L,\,Q_R, \\\,L_L,\,L_R}} \bar\psi \gamma_\mu X_\psi \psi  \\ \nonumber
		\ &-&c_{\tilde{W}}\WWd^a\tilde{W}^{a\mu\nu}\dfrac{a}{f_a}-c_{\tilde{B}} B_{\mu \nu} \tilde{B}^{\mu \nu} \dfrac{a}{f_a}.
\eea
Where $\WWd^a$ and $B_{\mu \nu}$ are the field strength tensors of $SU(2)_L$ and $U(1)_Y$, and $c_{\tilde{W}}$ and $c_{\tilde{B}}$ denote the corresponding coupling constants. And $X_\psi$ are hermitian matrices in flavour space.
After electroweak symmetry breaking, the Lagrangian can be redescribed as\cite{EFTandCollider}
\bea
\mathcal{L}_{\text {eff }}^{D \leq 5} &=& \dfrac{1}{2} (\partial_\mu a) (\partial^\mu a)- \dfrac{m_a^{2} a^2}{2}+ ia g_{a\psi} \sum_{\psi=Q,\,L}  m^{\text{diag}}_{\psi}\, \bar{\psi} \gamma_5 \psi , \\ \nonumber
\ &-&\dfrac{1}{4}g_{a\gamma\gamma}\, a\, F_{\mu\nu} \tilde{F}^{\mu\nu}-\dfrac{1}{4}g_{a\gamma Z}\,a\,F_{\mu\nu} \tilde{Z}^{\mu\nu}-\dfrac{1}{4} g_{aZZ}\,a\,Z_{\mu\nu} \tilde{Z}^{\mu\nu}-\dfrac{1}{4}g_{aWW}aW_{\mu\nu}\tilde{W}^{\mu\nu},
\eea
where
\bea
\begin{aligned}
	\centering
  \quad \quad &g_{a\gamma\gamma}=\frac{4}{f_{a}}(c_{\theta}^{2} c_{\tilde{B}} + s_{\theta}^{2} c_{\tilde{W}}) ,\\
 &g_{aZZ}=\frac{4}{f_{a}}(s_{\theta}^{2} c_{\tilde{B}}+c_{\theta}^{2}c_{\tilde{W}}), \\
 & g_{a\gamma Z}=\frac{4}{f_{a}}s_{2\theta}(c_{\tilde{W}}- c_{\tilde{B}}),\\
 & g_{aWW} = \frac{4 c_{\tilde{W}}}{f_{a}}.
  \end{aligned}
\eea
Here, all the couplings $g_{a\gamma\gamma}$, $g_{a\gamma Z}$, $g_{aZZ}$, $g_{aWW}$ and $g_{a\psi}$ are governed by the characteristic scale $f_a$. $F_{\mu\nu}$, $Z_{\mu\nu}$ and $W_{\mu\nu}$ are the photon, $Z$ boson and $W$ boson field strength tensors, respectively, and their duals are defined as $\tilde{X}^{\mu\nu}=\dfrac{1}{2}\varepsilon^{\mu\nu\alpha\beta}X_{\alpha\beta}$ with $\varepsilon^{0123}=+1$. The matrix $m^{diag}_{\psi}$ represents diagonalizable fermion mass matrix. Additionally, $s_{\theta}$ and $c_{\theta}$ are the sine and cosine of the weak mixing angle $\theta$, respectively.

The bounds on the couplings of ALPs to gluons, photons, fermions and massive gauge bosons have been extensively studied~\cite{EFTandCollider,ALPCollider,ALPMevto90,ALPCollider2,ALPee,ALPbosons,ALPtriboson,ALPFCNC,ALPEDM,Flavorcons,ALPFlavor,ALPLFV,ALPLFV2,ALPBelle,agg,nonresggF,nonresVBS}. Generally, the bounds depend on the ALP mass range considered. The most stringent limits on ALP couplings arise from cosmological and astrophysical bounds, which are valid for very light ALPs with masses below the MeV. For ALP with mass higher than MeV, the constrains are weaker. Significant constraints mainly come from CDF collaboration, LEP and LHC experiments for ALPs in the MeV to several tens GeV range~\cite{EFTandCollider,ALPCollider,ALPMevto90,ALPCollider2}. Especially, the LEP data on $Z\rightarrow \gamma \gamma$ decay provides significant constraints on $a\gamma \gamma$ and $a\gamma Z$ couplings in the range 1 MeV$<m_a<10$ GeV~\cite{ALPMevto90}. The couplings of ALPs to the massive gauge bosons are constrained mainly due to loop induced ALP decaying into two photons~\cite{EFTandCollider,ALPCollider2} and flavor-changing neutral
current meson decays~\cite{ALPFCNC}. Focusing on non-resonant ALP-mediated vector boson scattering processes, constraints on ALP-boson couplings are obtained from analysis of LHC RUN 2 data in a recent study~\cite{nonresVBS}, which include an upper bound $g_{aZZ}<2.84~\text{TeV}^{-1}$ . Other limits come from LHC mono-$W$, mono-$Z$ searchs~\cite{EFTandCollider} and non-resonant searches of two bosons via the gluon-gluon fusion~\cite{nonresggF}. Furthermore, a nonzero $a\gamma Z$ coupling induces a exotic decay $Z\rightarrow \gamma a$. The $Z$-boson total width measurement performed at LEP~\cite{LEP} can put a severe constraint on the coupling $g_{a\gamma Z}$. Neglecting the ALP mass, it is found that~\cite{EFTandCollider} :
\beq
g_{a\gamma Z}<1.8~\text{TeV}^{-1}~~(95\% C.L.)
\eeq
It is notably that this bound gets weaker as $m_a$ increases. For example, it is found that $g_{a\gamma Z}< 6.8~\text{TeV}^{-1}~$(95\% C.L.) when $m_a= 70$ GeV. Focusing on lepton sector, interesting bounds on ALP-fermion interactions can be obtained from several experimental data including the search for light $Z'$ bosons performed by BaBar \cite{BaBar}, and the measurement of lepton anomalous magnetic moment $a_{l}$~\cite{mug2, eg2}. Actually, the most stringent constraint on ALP-fermion coupling $g_{a\psi}$ for ALPs below 10 GeV comes from the Bean Dump experiments (CHARM) by testing rare meson decays\cite{Flavorcons,CHARM:1985anb}, as pointed out in Ref.\cite{EFTandCollider}:
\beq
 g_{a\psi} <3.4 \cdot 10^{-5} - 2.9\cdot 10^{-3}~\text{TeV}^{-1}~~(90\% C.L.)~~~\text{for}~~ 1 \text{MeV}\lesssim m_a  \lesssim 3 \text{GeV}.
\eeq

In order to accommodate to experimental constraints, the mass of ALP is assumed in the range 5 GeV$<m_a<70$ GeV and the values of the couplings 
$g_{aZZ}$ and $g_{a\gamma Z}$\footnote{For ALPs with masses approaching $m_Z$, a small parameter region with larger couplings is also shown in the FIG~\ref{fig:curve}.} are assumed to be less than 2.8 TeV$^{-1}$ and 1.8 TeV$^{-1}$, respectively.

\section{Searching for ALP via exotic $Z$ decay at future $Z$ factories}

\begin{figure}[htb]
\begin{center}
\includegraphics [scale=0.65] {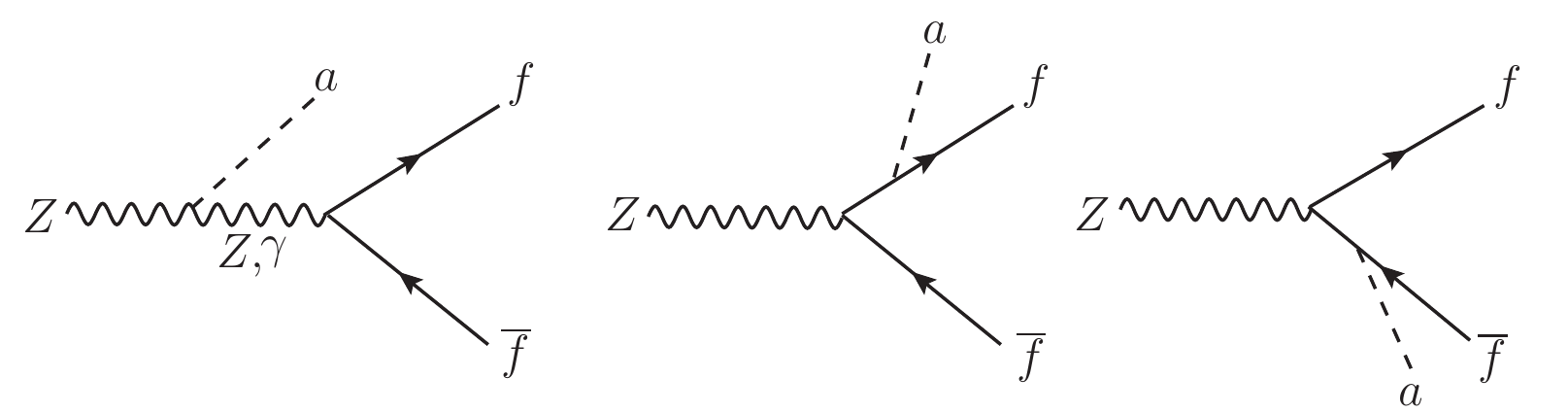}
\caption{The Feynman diagrams for exotic $Z$ decay $Z \rightarrow a f\bar{f}$ where $f$ represents fermions except top quark. }
\label{fig:feyaxll}
\end{center}
\end{figure}

The CEPC~\cite{CEPCPCDR,CEPCCDR} and FCC-ee~\cite{FCCee1,FCCeeCDR,FCCCDR} are the two options for next generation $e^{-}e^{+}$ colliders, which will first operate around the $Z$-pole for 2-4 years and thus called $Z$ factories. More than $10^{12}$ $Z$ boson events will be produced during this period. Such large number of $Z$ boson events at the CEPC and FCC-ee would open up a new avenue to directly probe the light particles produced from rare $Z$ decays~\cite{DMwithZ,Zhang1,Chang}. Previous studies also found that the future $e^{-}e^{+}$ colliders have the potential to probe ALPs via ALP-strahlung~\cite{ALPee}, light-by-light scattering~\cite{Zhang1} and vector boson fusion processes~\cite{Zhang2}. For an ALP with mass $m_a< m_Z$, there exits  rare decay channel $Z \rightarrow af\bar{f}$, where $f$ can be neutrinos, charged leptons or quarks. The relevant Feynman diagrams are shown in Fig.~\ref{fig:feyaxll}. It is expected that future $Z$ factories have the potential to directly detect the ALP in these channels. In the absence of the $a\nu\bar{\nu}$ interaction, the decay $Z \rightarrow a\nu \bar{\nu}$ is only induced via ALP emitted from $Z$ boson. The relevant measurements of this channel could be used to extract the $aZZ$ coupling directly.
Unlike the decay $Z\rightarrow a \nu\bar{\nu}$, all of the couplings $aZZ$, $a\gamma Z$ and $af\bar{f}$ can contribute to the decay $Z\rightarrow a f\bar{f}$ with $f$ being charged leptons or quarks. However, this channel is more sensitive to the coupling $a\gamma Z$. This is partially because the coupling of ALP to fermions is suppressed by the fermion mass. Additionally, the Feynman amplitude involving the parameter $g_{aZZ}$ is suppressed in the $m_a$ range described for details in section III C.

In this paper, we will focus on the decay channels  $a\rightarrow \mu^+ \mu^-$ and  $a\rightarrow b\bar{b}$ due to relative large Yukawa couplings and good performance of muon identification and $b$-tagging. Four types of exotic $Z$-decay signals $Z \rightarrow  \mu^+ \mu^- \slashed E$, $ b b\slashed E$, $ e^+ e^- \mu^+ \mu^- $ and $ e^+ e^- b b $ are studied, respectively in the following subsections.

In the simulation, we use FeynRules~\cite{FR} to generate the model file for the effective Lagrangian. The signal and background events are generated by MadGraph5-aMC@NLO~\cite{MG5} and then fed to PYTHIA 8~\cite{PYTHIA} to simulate initial state radiation, final state radiation and hadronization. Finally, a fast detector simulation is carried out by Delphes 3 and the FCC-ee delphes card is used~\cite{DELPHES}. The package Madanalysis 5~\cite{MA5} is adopted for event analysis. The collision energy $E_{CM}=91.2$ GeV is chosen. To simulate the detector acceptance and preselect events, we employ the basic cuts for signals and backgrounds. We require that the transverse momentum $P_T$ of lepton and jet (including $b$ jet) is larger than 10 GeV. For the events with missing energy, we require $\slashed E > 10$ GeV as well\footnote{Unlike the missing transverse energy $\slashed E_T$ at the large hadron collider, the missing energy $\slashed E$ is reconstructed here due to the full 4-momentum information of initial state and clean environment at electron colliders.}.  We also require that the pseudorapidity of all visible particles satisfy $|\eta(l,j)| <2.5$ . In addition, the separation requirements are $\theta_{ij}(l,l)>0.2$, $\theta_{ij}(j,j(l))>0.4$ for leptons and jets, respectively.
The basic cuts are summarized as:
\beq
P_T(l,j)>10 \text{GeV}, \ \slashed E > 10\text{GeV} ,\ |\eta(l,j)|<2.5, \ \theta_{ij}(l,l)>0.2,\ \theta_{ij}(j,j)>0.4,
\eeq
where $l=e,\mu$ and $j$ includes both the light flavor jets and $b$ jets.


\subsection{ $Z \rightarrow  \mu^+ \mu^- \slashed E$  }

\begin{figure}[H]
\begin{center}
\subfigure[]
{
\label{fig:2a}
\includegraphics [scale=0.38] {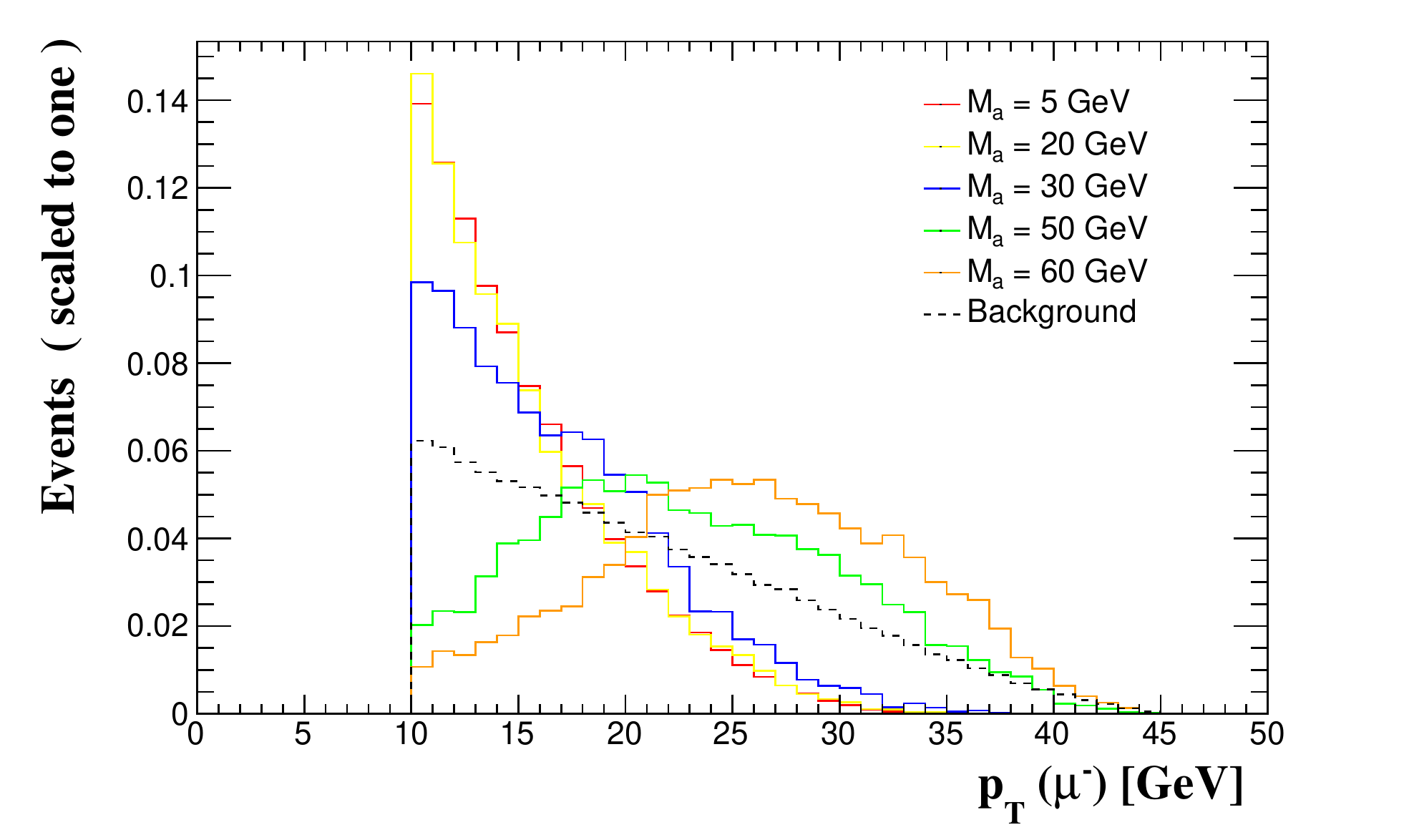}}
\subfigure[]
{
\label{fig:2b}
\includegraphics [scale=0.38] {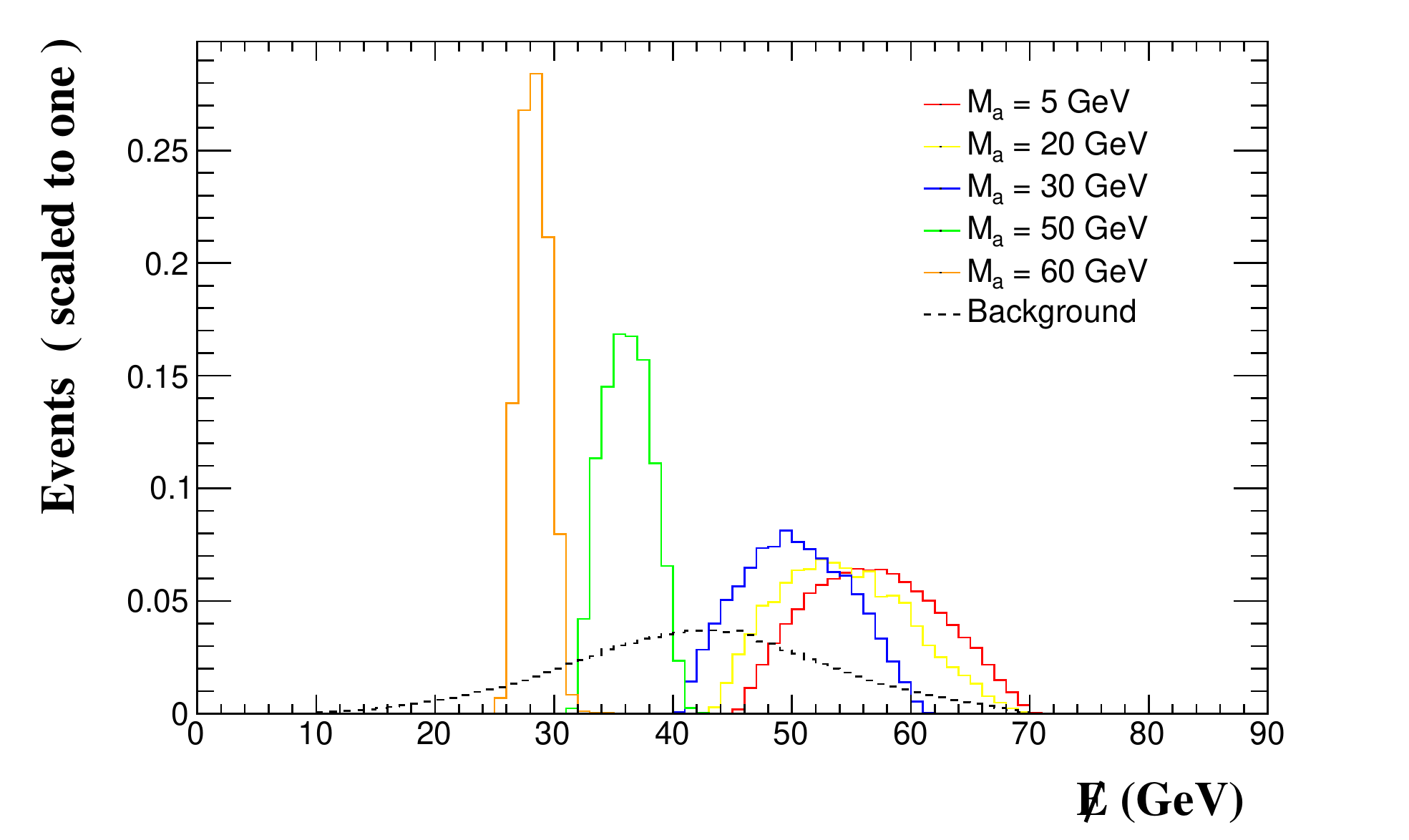}}
\caption{The normalized distributions for transverse momentum $P_T(\mu^-)$ and missing energy $\missE$ for the signal $\mu^+ \mu^- \slashed E$ and the corresponding SM background.}
\label{fig:DismumuE}%
\end{center}
\end{figure}

Let us begin with the exotic decay $Z\rightarrow \nu \bar{\nu} a $ followed by $a\rightarrow \mu^+ \mu^-$.
For the signal $\mu^+\mu^- \slashed E$, the background comes from $\mu^+ \mu^- \bar{\nu} \nu$ mediated by off-shell gauge boson $\gamma^*$, $Z^*$ and $W^*$. The normalized distributions for transverse momentum of $\mu^-$ and missing energy for both the signal and background events are shown in Fig.~\ref{fig:DismumuE}. For the signal events, the heavier the mass of ALPs, the harder of muons tend to. A large fraction of the SM background events have softer muons, which have been filtered out by the basic cuts. When the ALP is light, the missing energy is large and it has a wide distribution due to relative large phase space. Considering only two visible particles in the final states, the invariant mass cut of a pair of muons $m_{\mu^+\mu^-}$ is adopted as  $|m_{\mu^+\mu^-}-m_a|<$3 GeV (Cut 1-A).

\begin{table}[tbh]
\begin{center}%
\begin{tabular}
[c]{|c|c|c|c|c|c|}\hline
\multirow{2}{*}{Cuts}    & \multicolumn{5}{c|}{ Cross sections for signal(background) (fb) }   \\
\cline{2-6}
	       &$m_a$ = 5GeV      &$m_a$ = 10GeV   & $m_a$ = 30GeV       & $m_a$= 50GeV      &  $m_a$= 60GeV   \\ \hline
	Basic cuts      &  0.3406(0.2602)  & 0.3177(0.2602) &  0.2368(0.2602)    & 0.0358(0.2602)   & 0.0062(0.2602)   \\ 
	Cut 1-A         &  0.3404(0.0059)  & 0.3175(0.0072)&  0.2335(0.0300)    & 0.0343(0.0342)   & 0.0058(0.0195)    \\ 
	$S/\sqrt{S+B}$  &     18.29        & 17.62   &     14.38         &   5.59   &   1.16  \\ \hline

\end{tabular}
\end{center}
\caption{Signal and background cross sections for the signal $\mu^+\mu^- \missE $ after cuts applied for $g_{aZZ}=0.5~\text{TeV}^{-1}$. The statistical significance(SS) is calculated for an integrated luminosity of 1$\text{ab}^{-1}$.}%
\label{table:mumumissE}%
\end{table}
The comparison of the cut efficiencies between the signal and background is shown in TABLE~\ref{table:mumumissE}.
Here, $g_{aZZ}=0.5$ TeV$^{-1}$ are assumed and five cases $m_{a}= 5$, $10$, $ 30$, $ 50 $ and $60 $ GeV are presented respectively for paradigm. Generally, if both the decays $a\rightarrow \mu^+\mu^-$ and $a\rightarrow b\bar{b}$ are opened, the branching ratio Br($a\rightarrow \mu^+ \mu^-)$ is suppressed by its relative small Yukawa coupling. However, there exist some lepton-philic ALP models with rich phenomenology~\cite{LP-ALP}. For simplicity, we have taken Br($a\rightarrow \mu^+ \mu^-)=100\% $. After imposing mass window cut, the background has been significantly rejected.
Taking an integrated luminosity of $1\text{ab}^{-1}$, a significance $SS = S/\sqrt{S+B}=18.29$ (5.59) for $m_{a}= 5$ GeV (50 GeV) can be achieved. For heavier ALPs, a larger integrated luminosity is needed for exclusion or discovery.

\subsection{$Z \rightarrow  b b \slashed E$}

If $m_a > 2m_b$, the dominant decay channel of ALP is $a \rightarrow b \bar{b}$ due to large Yukawa coupling of bottom quarks. In this subsection, we discuss the signal $b b \slashed E$. For this signal, the SM backgrounds are dominantly from $b \bar{b} \nu \bar{\nu}$ and  $j j \nu \bar{\nu}$ mediated by off-shell gauge boson $\gamma^*$, $Z^*$ and $W^*$. The kinematic distributions for the signal and background are similar to the case of $\mu^+\mu^- \slashed E$. So we employ a similar cut in the $bb \slashed E$ case. While the reconstructed $b$ pairs is requred in the mass window $|m_{bb}-m_a|<$5 GeV (Cut1-B).

\begin{table}[htb]
\begin{center}%
\begin{tabular}
[c]{|c|c|c|c|c|c|}\hline
\multirow{2}{*}{Cuts}    & \multicolumn{5}{c|}{ Cross sections for signal(background) (fb) }   \\
\cline{2-6}
               &$m_a$ = 15GeV        & $m_a$ = 30GeV      & $m_a$= 40GeV    & $m_a$= 50GeV      & $m_a$=60GeV  \\ \hline
Basic cuts     & 0.0460(0.7172)    &0.03345(0.7172)   & 0.0284(0.7172)      & 0.0142(0.7172)    & 0.0134(0.7172)   \\ 
Cut 1-B        & 0.0449(0.0126)    &0.0279(0.08387)    & 0.0199(0.1832)      & 0.0078(0.2115)    & 0.0015(0.1364)   \\ 
$S/\sqrt{S+B}$   & 5.92            & 2.64             & 1.39                & 0.53              & 0.12    \\ \hline

\end{tabular}
\end{center}
\caption{Same as TABLE I but for the signal $ bb \missE $.}%
\label{table:bbmissE}%
\end{table}

The cut efficiencies are shown in TABLE~\ref{table:bbmissE}. In the basic cuts, we include the requirement of two tagged $b$ jets. We assume a $b$-tagging efficiency of $80\%$ and a mis-tagging rate of $10\%$ for $c$ jet and $1\%$ from light flavor $(u, d, s$ or $g)$ jet. The SM background $j j \nu \bar{\nu}$ is significantly suppreesed by two $b$ taggings.  After these cuts, the SM background are reduced to a controlled level and a $5.92\sigma$ ($2.64\sigma$) significance can be achieved when $m_a =15 \textrm{GeV}$ (30 GeV). For heavier ALPs, a higher luminosity is needed for detecting.

\subsection{ $Z \rightarrow e^+ e^- \mu^+ \mu^- $ }

\begin{figure}[htb]
\begin{center}
\subfigure[]
{
\includegraphics [scale=0.38] {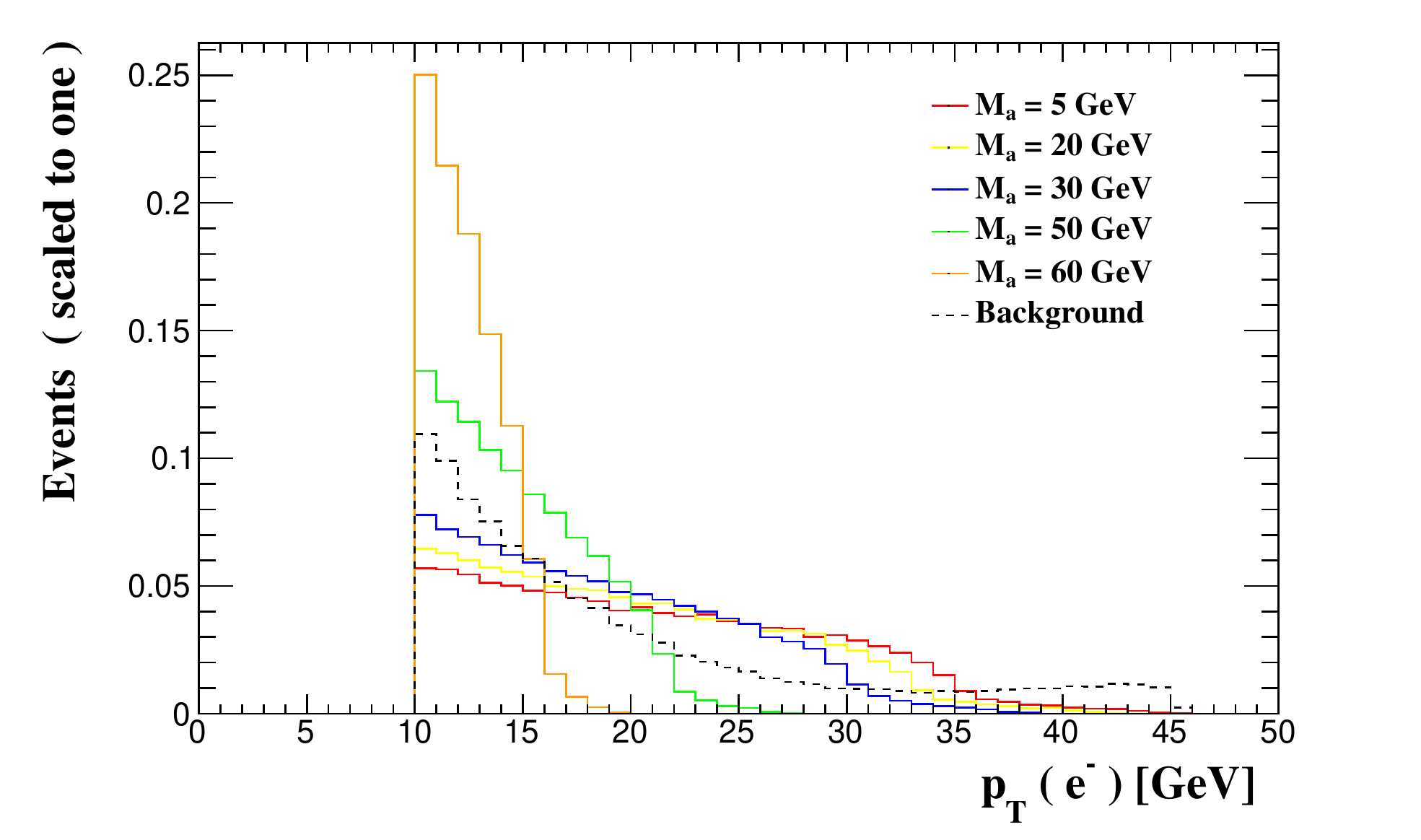}}
\subfigure[]
{
\includegraphics [scale=0.38] {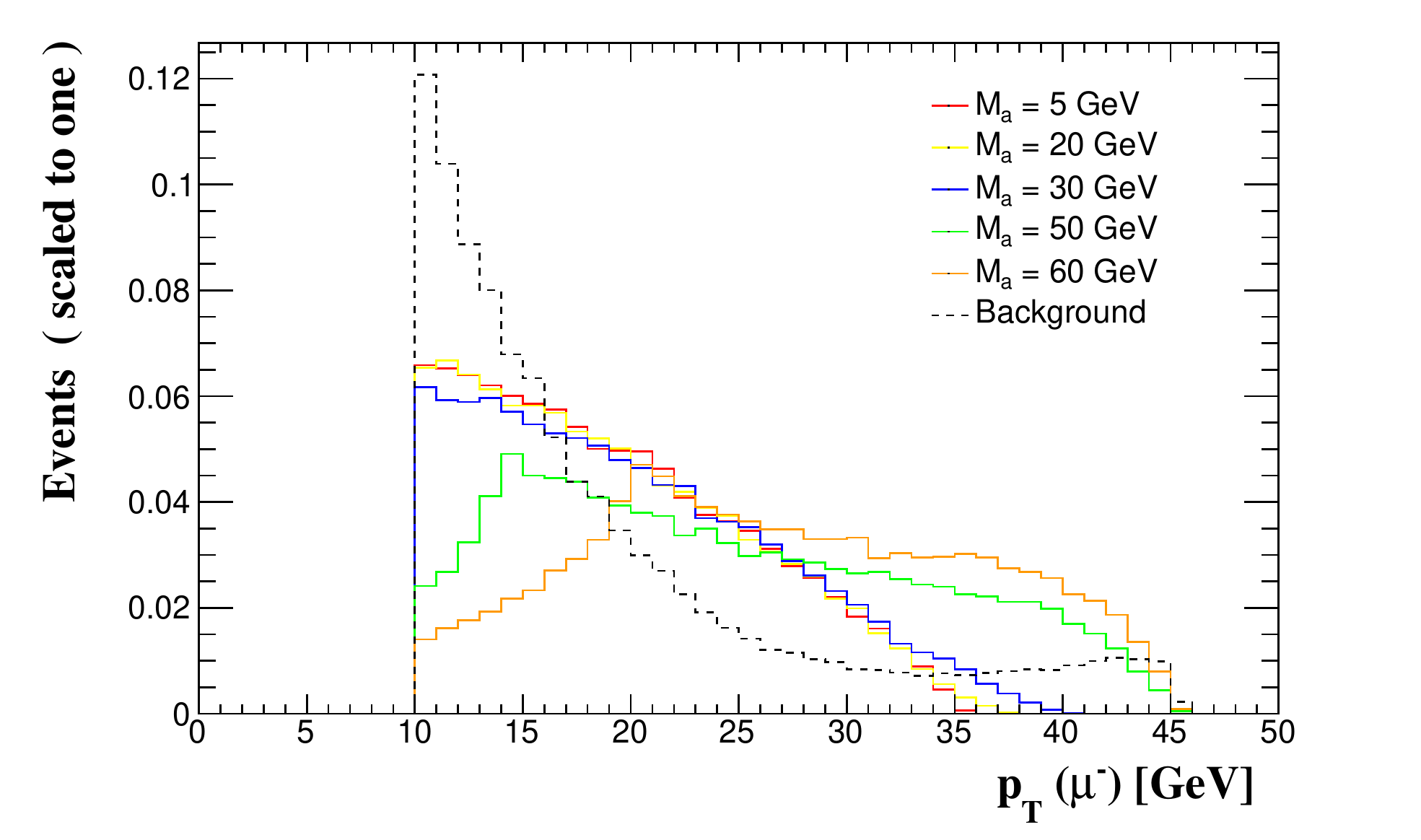}}
\subfigure[]
{
\includegraphics [scale=0.38] {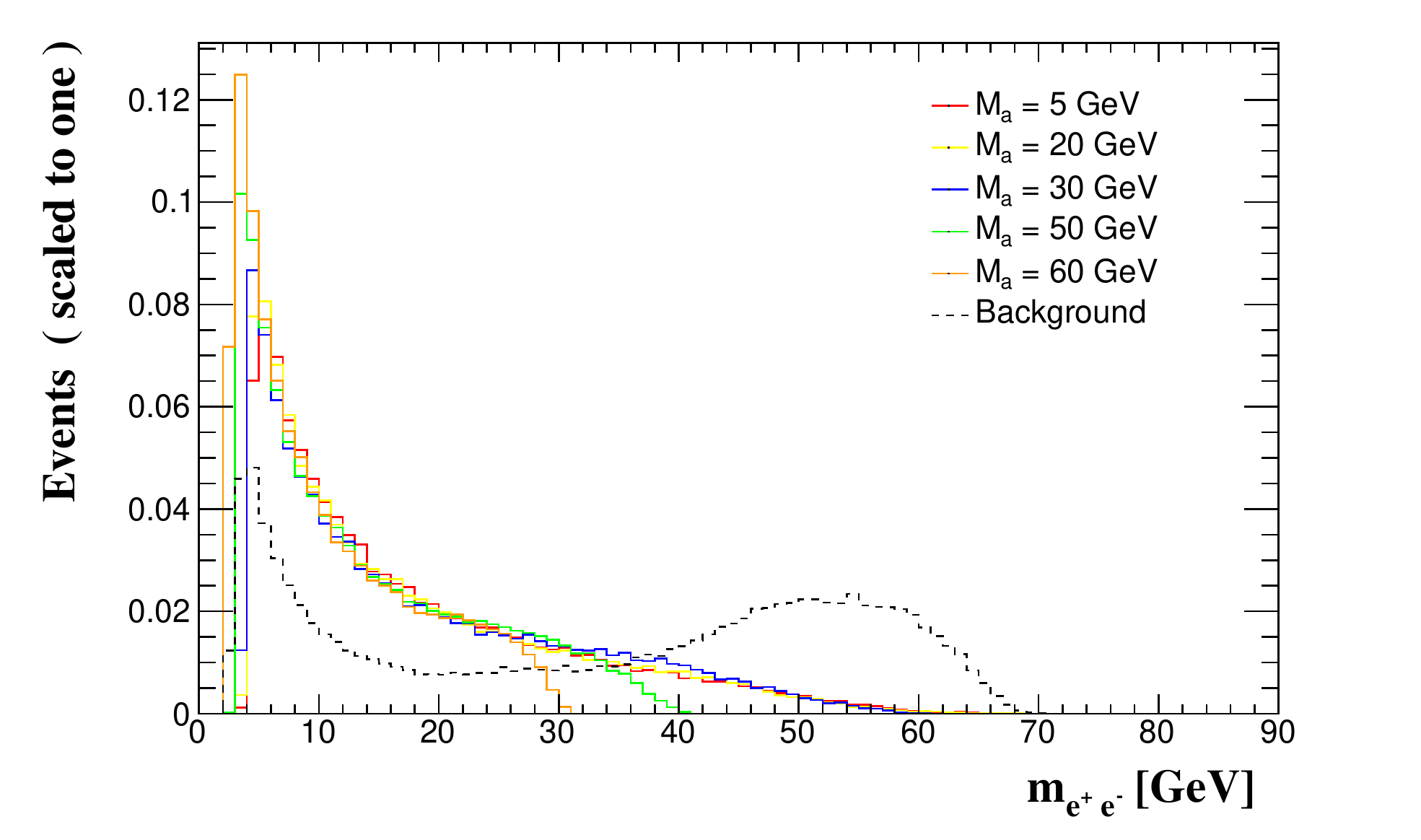}}
\subfigure[]
{
\includegraphics [scale=0.38] {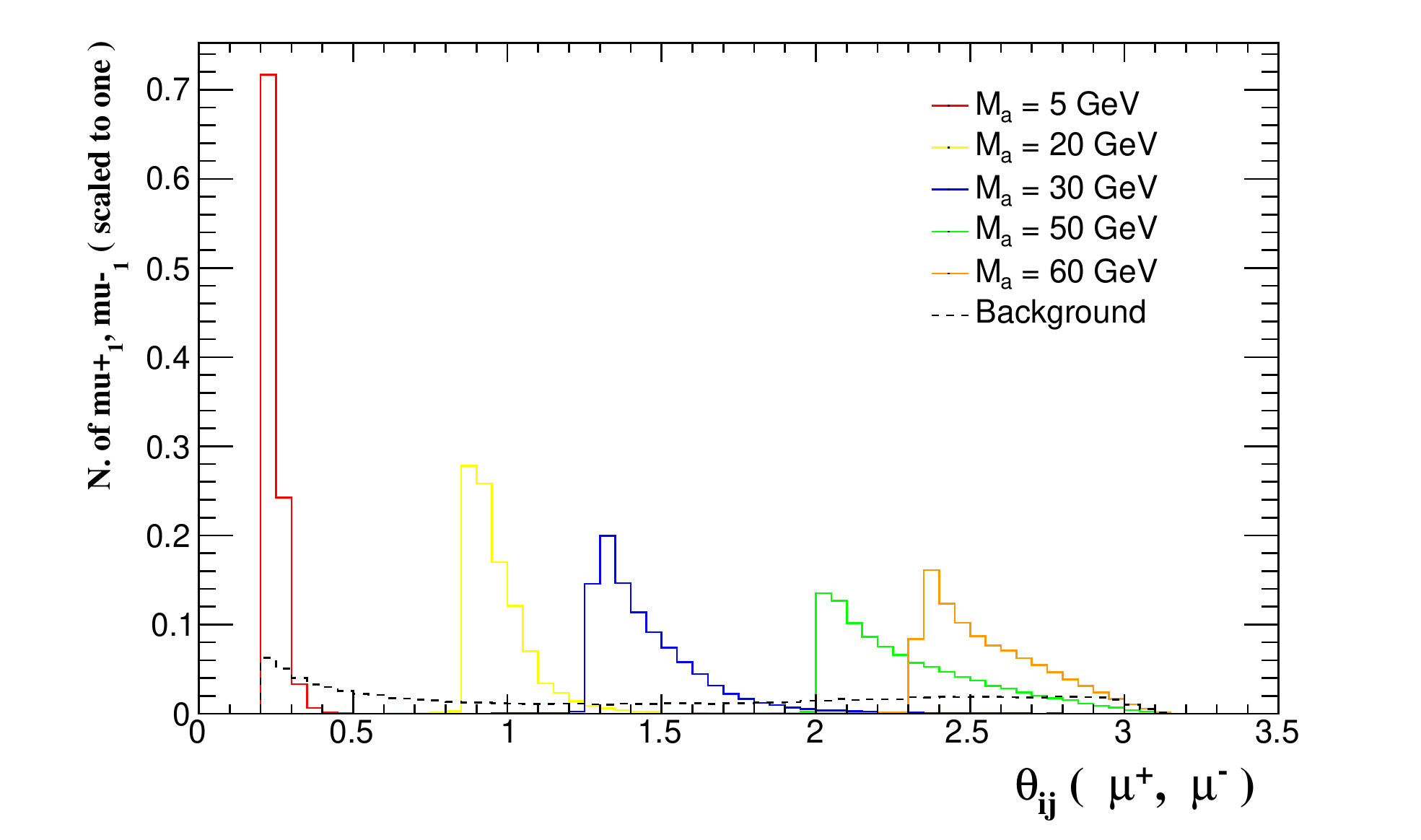}}
\caption{The normalized event distributions for kinematic variables for $P_T(e^-)$, $P_T(\mu^-)$, $m_{e^+, e^-} $ and $\theta_{ij}(\mu^+,\mu^-)$ for signal $e^+e^-\mu^{+}\mu^{-}$ and the background. }
\label{fig:Diseemumu}
\end{center}
\end{figure}
In this subsection, we explore the prospects for detecting ALP particles via the decay $Z \rightarrow e^+ e^- \mu^+ \mu^-$.
The dominant SM background for the final state of $e^+e^-\mu^+\mu^-$ is mainly coming from $Z^*Z^*$, $Z^*\gamma^*$, $\gamma^*\gamma^*$ production.
Normalized distributions for transverse momentum of electon and muon, as well as invariant mass of electron pairs for signal and background events are shown in Fig.~\ref{fig:Diseemumu}. For the signal events, the $P_T$ of $\mu^-$ tends to have a wide distribution due to the kinematics of three-body decay. While for the background events, the muons are always soft. One can also see that the pair of muons from light ALPs tends to collimated and that from heavy ALPs tends to back-to-back. For the background, there is a relative large fraction of events have the pair of muons with a small separation which coming from a boosted $\gamma^*$.

According to three-body decay kinematics in $Z\rightarrow e^+e^-a $ channel, the reconstructed invariant mass of electron pair $m(e^+e^-)$ in the $Z$ rest frame can be described as:
\beq
m^2_{e^+e^-}=[P_Z-P_a]^2=m^2_Z+m^2_a-2m_ZE_a.
\eeq
Generally, the distribution of $m_{e^+e^-}$ possesses a maximum value at $m_{e^+e^-}=m_Z-m_a$ as shown in Fig.~\ref{fig:Diseemumu}(c). For a light ALP, a nonzero momentum is needed to generate two muons passing the basic $P_T$ cut, which shifts the maximum. The $m_{e^+e^-}$ distribution is incline to have a small value. So the amplitude from $g_{aZZ}$ contribution is approximately suppressed by $1/m^2_Z$. In addition, the coupling of ALP to fermions is suppressed by small mass factor. Hence, the signals $e^+ e^- \mu^+ \mu^-$ and $ e^+ e^- b b$ are more sensitive to the coupling $g_{a\gamma Z}$. Whereas, for the background, there is a large fraction of events with a harder $e^+e^-$ mass spectrum which originates from $Z^*$. In this subsection and following subsection, we will take $g_{aZZ}=0$ for simplicity.

\begin{table}[htb]
\begin{center}%
\begin{tabular}
[c]{|c|c|c|c|c|c|}\hline
\multirow{2}{*}{Cuts}    & \multicolumn{5}{c|}{ Cross sections for signal(background) (fb) }   \\
\cline{2-6}
               &$m_a$ = 5GeV         & $m_a$ = 10GeV    & $m_a$= 30GeV      & $m_a$= 50GeV   &  $m_a$= 60GeV   \\ \hline
Basic cuts     &  1.5314(8.0284)     &  1.4735(8.0284)  & 1.1559(8.0284)  & 0.4615(8.0284)  & 0.1067(8.0284)  \\ 
Cut 1-C        &  1.2659(3.4300)     &  1.2231(3.4300)  & 0.9318(3.4300)  & 0.4258(3.4300)   & 0.1065(3.4300)   \\ 
Cut 2-C        &  1.2659(0.1743)     &  1.2215(0.1623)  & 0.9143(0.1689)  & 0.4066(0.6499)   & 0.1005(0.5635)  \\
$S/\sqrt{S+B}$ &  33.36              &     32.83       &   27.78        &   12.51          &  3.90   \\ \hline
\end{tabular}
\end{center}
\caption{Same as TABLE I but for the signal $ e^+ e^-\mu^+\mu^- $ and $g_{a\gamma Z}=0.5~\text{TeV}^{-1}$.}%
\label{table:eemumu}
\end{table}

According to above analysis, we employ a cut on $m_{e^+e^-}$ to suppress the background. Furthermore, a mass window cut on the reconstructed muon pair with hypothesis $m_a$ mass can improve the significance. The optimized cuts are summarized as follows:
\begin{enumerate}
\item Cut1-C: the invariant mass of electron pairs satisfies $m_{e^+ e^-}< 30$ GeV.
\item Cut2-C: the mass window cut $|m_{\mu^+\mu^-}-m_a|<3$ GeV.
\end{enumerate}

We take basic cuts and optimized cuts step by step and show the statistical significance for an integrated luminosity of 1ab$^{-1}$ in TABLE~\ref{table:eemumu}.
It can be seen that large values of $SS$ can be achieved in wide range of the parameter space.


\subsection{ $Z \rightarrow e^+ e^- b b $ }

In this subsection, we focus on the decay channel $Z \rightarrow e^+ e^- b b$. The relevant SM backgrounds are $e^+e ^-jj$  and $e^+e ^-bb$. The event topologies for the signal are the same as that of $ Z \rightarrow e^+ e^- \mu^+ \mu^-$, so we propose a set of similar cuts as above. Besides basic cuts including two $b$-taggings requirement, the optimized cuts are summarized as: $m_{e^+ e^-}< 30$ GeV (Cut 1-D) and $|m_{bb}-m_a|<5$ GeV (Cut 2-D).

\begin{table}[htb]
\begin{center}%
\begin{tabular}
[c]{|c|c|c|c|c|c|}\hline
\multirow{2}{*}{Cuts}    & \multicolumn{5}{c|}{ Cross sections for signal(background) (fb) }   \\
\cline{2-6}
               &$m_a$ = 15GeV       & $m_a$ = 30GeV     & $m_a$= 40GeV   & $m_a$= 50GeV   &  $m_a$= 60GeV   \\ \hline
Basic cuts     & 0.2058(2.7076)     &0.2497(2.7076)		&0.1874(2.7076) & 0.1191(2.7076)  & 0.0354(2.7076)                 \\ 
Cut 1-D        & 0.1711(1.8332)     &0.2251(1.8332)     &0.1658(1.8332) & 0.1103(1.8332)  & 0.0353(1.8332)		    \\ 
Cut 2-D        & 0.1694(0.1417)     &0.1763(0.2081)     &0.1106(0.3301) & 0.0617(0.5102)  & 0.0160(0.4212)  \\
$S/\sqrt{S+B}$ & 9.61               &8.99               &5.27           & 2.58            & 0.76              \\ \hline
\end{tabular}
\end{center}
\caption{Same as TABLE I but for the signal $ e^+ e^- b b $ and $g_{a\gamma Z}=0.5~\text{TeV}^{-1}$.}%
\label{table:eebb}%
\end{table}

As shown in the TABLE \ref{table:eebb}, the backgrounds are suppressed by selection cuts. After step-by-step cuts, a significance larger than $5\sigma$ can be achieved for $m_{a}<40 \textrm{GeV}$ with an integrated luminosity of 1$\text{ab}^{-1}$. For a heavier ALP, the production rate is suppressed by phase space and hence it has a small significance.

In this section, we have explored four types of signals $\mu^+\mu^-\slashed E$,  $b b\slashed E$, $ e^+ e^- \mu^+ \mu^-$ and $e^+ e^- b b$ at future $Z$ factories. It is notably that the signals $\mu^+\mu^-\slashed E$ and $ b\bar{b}\slashed E$ can be used to extract the coupling $g_{aZZ}$, whereas the signals $e^+ e^- \mu^+ \mu^-$ and $e^+ e^- b b$ are sensitive to the coupling $g_{a\gamma Z}$.

\begin{figure}[htb]
\begin{center}
\subfigure[]
{
\label{fig:curve-a}
\includegraphics [scale=0.26] {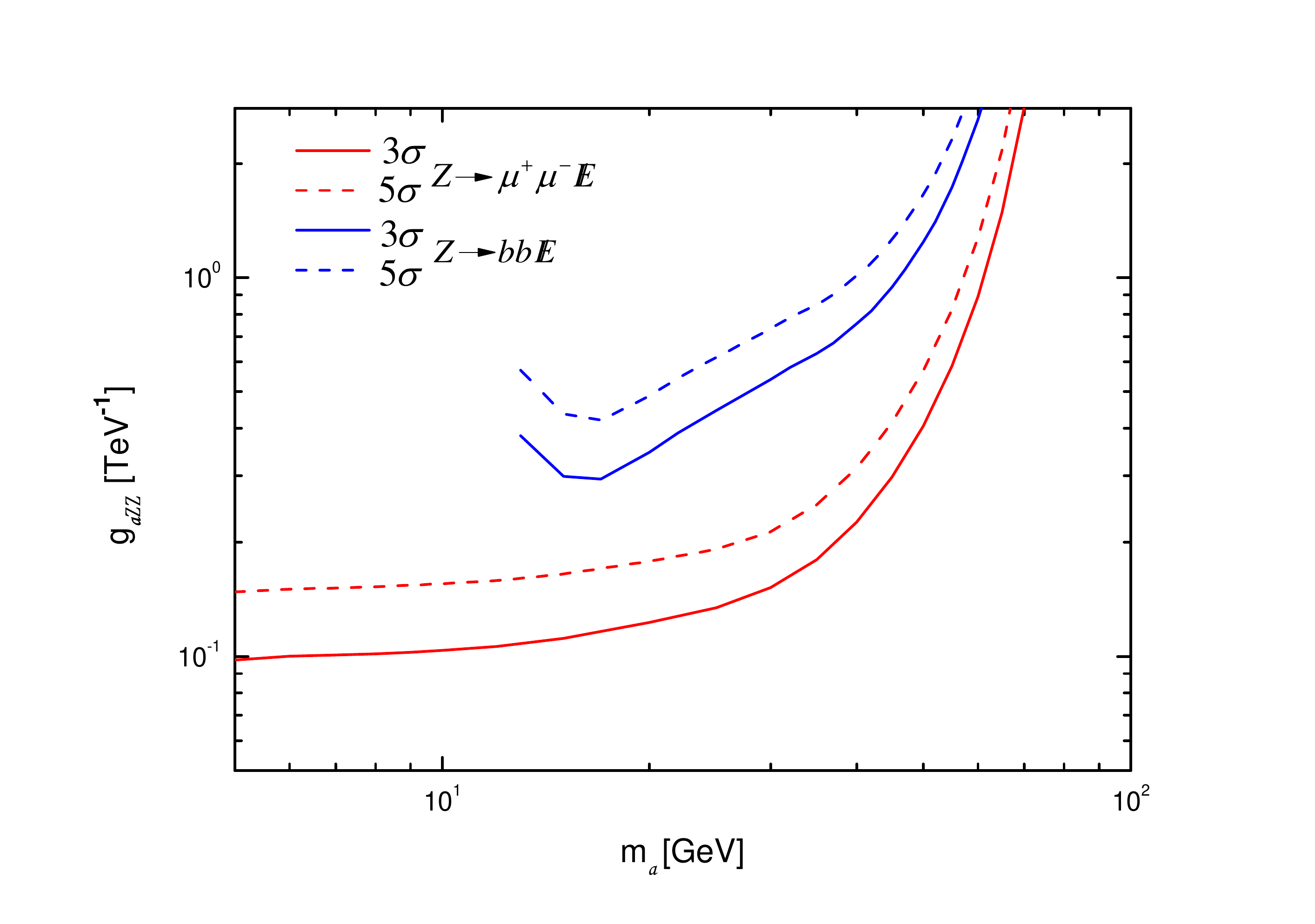}}
\subfigure[]
{
\label{fig:curve-b}
\includegraphics [scale=0.26] {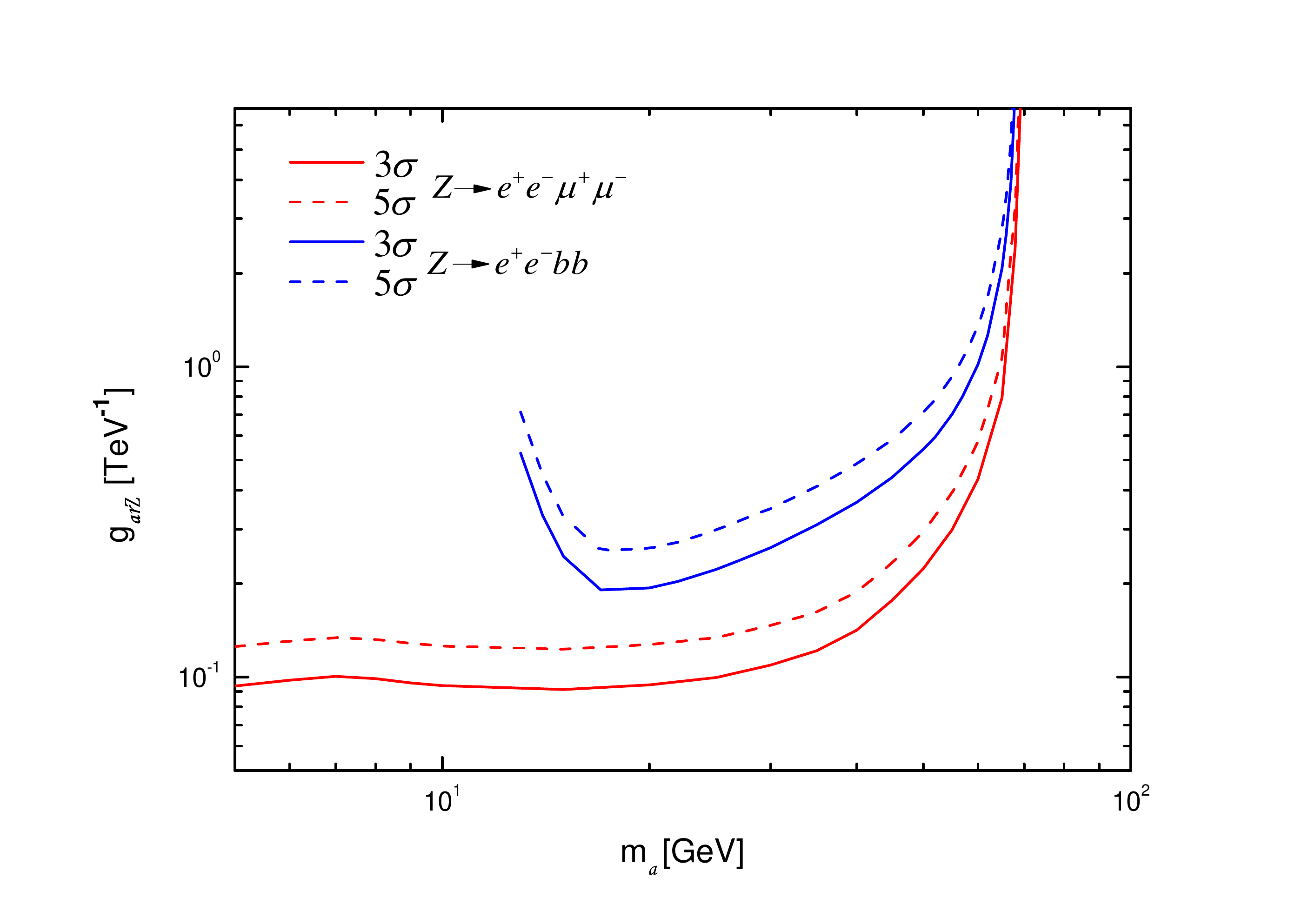}}
\caption{$3\sigma$ and $5\sigma$ discovery curves for the $Z$ factory with 1$\text{ab}^{-1}$ integrated luminosity
in the planes $(m_a,g_{aZZ})$ and $(m_a,g_{a\gamma Z})$, respectivly. Here, the red (blue) curves are for the signal $\mu^+\mu^- \slashed E$ ($bb \slashed E$)(left) and the red (blue) curves are for the signal $e^+e^-\mu^+\mu^-$ ($e^+e^-bb$)(right).}
\label{fig:curve}
\end{center}
\end{figure}

In Fig.~\ref{fig:curve}, we plot the $3\sigma$ and $5\sigma$ curves for the $Z$ factory with 1$\text{ab}^{-1}$ integrated luminosity in the planes $(m_a,g_{aZZ})$ and $(m_a,g_{a\gamma Z})$, respectivly.
As shown in Fig~\ref{fig:curve}, the future $Z$ factories have the potential for discovering ALPs in most of the mass range considered in this paper.
For the ALP mass $m_a$ approaching 70 GeV, it becomes tough to detect ALPs, especially for the signals $bb \slashed E$ and $e^+e^-bb$.

For 5GeV$\leq m_a \leq 20$GeV, the $\mu^+\mu^- \slashed E$ and $e^+e^-\mu^+\mu^-$ signals can probe the couplings $g_{aZZ}$ and $g_{a\gamma Z}$ down to 0.1 TeV$^{-1}$. The most sensitive area for the $b b \slashed E$ signal is around $m_a=15$ GeV. For ALPs with mass smaller than 13 GeV, it is difficult to select two separated tagged $b$-jet in the detector. Developing jet substructure techniques or optimized $b$-tagging algorithms could be helpful to improve the significance in this case, which is beyond the scope of this paper. For the signal $e^+e^-b b $, it is similar to that for the signal $b b \slashed E$.

\section{CONCLUSIONs and discusions}
Many extensions of the SM feature one or several spontaneously broken global $U(1)$ symmetries, thus predicting the existence of ALPs. The future $Z$ factories provide a clean environment and high luminosity to detect ALP via exotic $Z$ decays. In this paper, we have studied the possibility of probing ALP particles in future $Z$ factories. Focusing on the signals $\mu^+\mu^-\slashed E$,  $b b\slashed E$, $ e^+ e^- \mu^+ \mu^-$ and $e^+ e^- b b$, we have found that these four types of signals are promising to detect ALPs in large range of the parameter space.

\begin{figure}[htb]
\begin{center}
\subfigure[]
{
\label{fig:exclusion-gaxzz}
\includegraphics [scale=0.26]{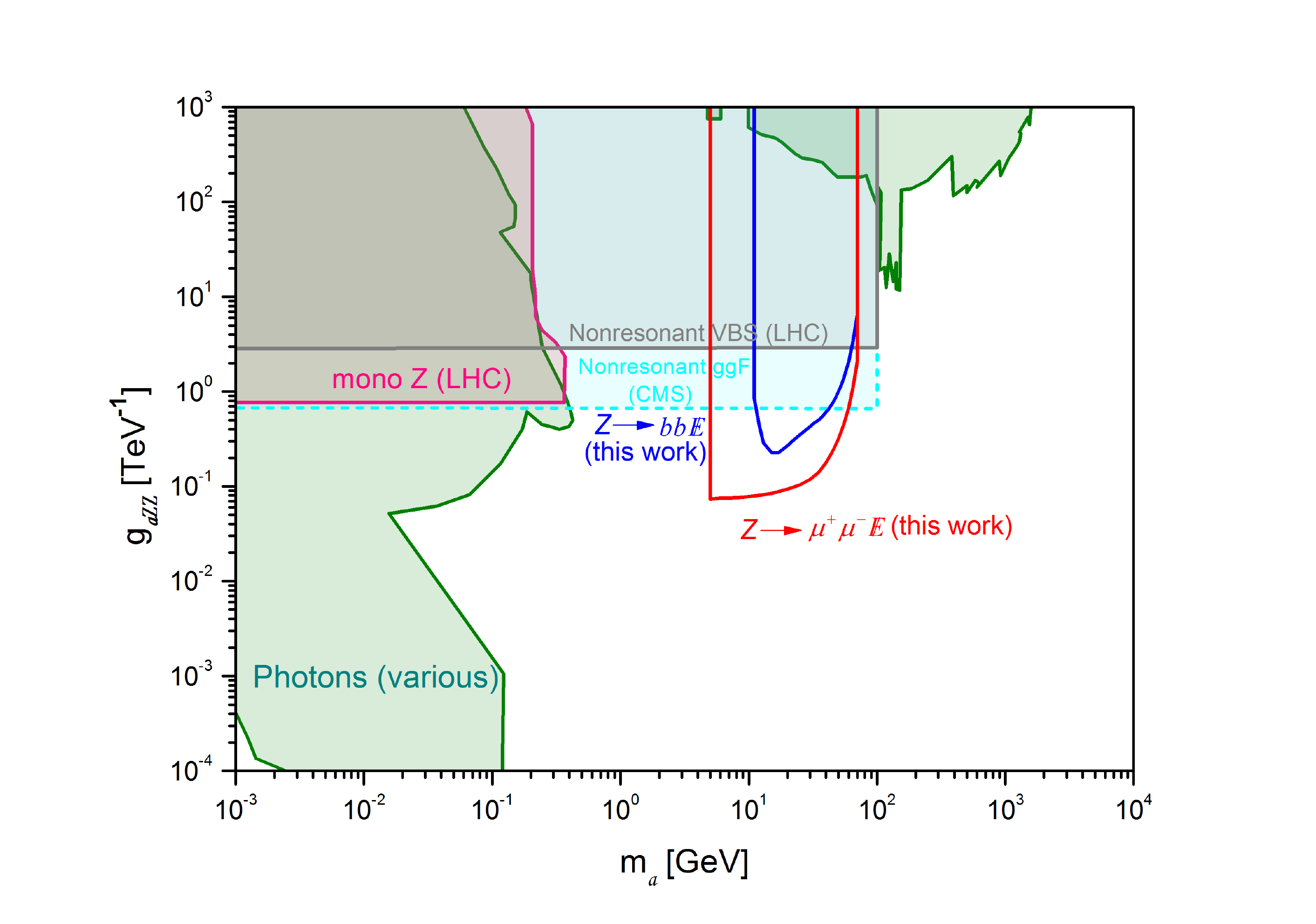}}
\subfigure[]
{
\label{fig:exclusion-gaxaz}
\includegraphics [scale=0.26]{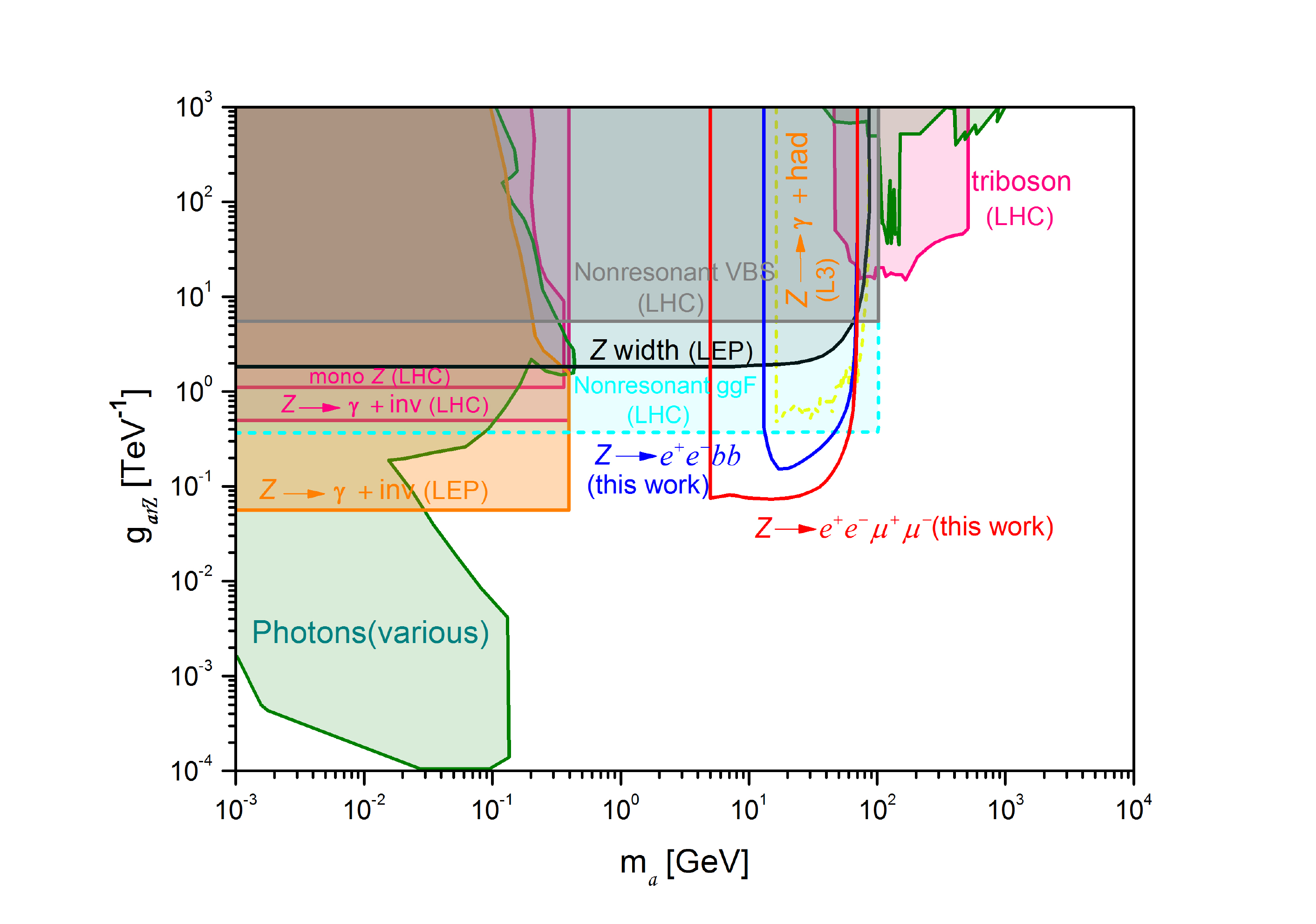} }
\caption{Sensitivity bounds on $g_{aZZ}$ (left) and $g_{a\gamma Z}$ (right) at 95\% C.L. from exotic $Z$ decays and other current exclusion regions. Sensitivity bounds derived in this work are labeled as corresponding signals and shown in red ($\mu^+\mu^-$ in final states) and blue ($bb$ in final states).}
\label{fig:exclusion}
\end{center}
\end{figure}

In Fig.~\ref{fig:exclusion}, we present the sensitivity bounds on the couplings $g_{aZZ}$ and $g_{a\gamma Z}$ obtained in this paper and the constraints from other studies. Most of the constraints are taken from Refs.~\cite{ALPbosons,nonresggF}. The constraints labelled ``Photons" are based on
the constraints on $g_{a\gamma \gamma}$ from supernova SN1987a observations, beam dump experiments, as well as LHC results. Due to radiative corrections of the ALP-boson couplings to the ALP-photon couplings, these results can be translated into constraints on $g_{aZZ}$ and $g_{a\gamma Z}$~\cite{ALPbosons}. For light ALPs, the mono-$Z$ and $\gamma+inv$ searches at colliders can put strong constraints on the ALP-boson couplings~\cite{EFTandCollider,monogammaATLAS}. For large ALP masses $m_a > m_Z$, Ref.~\cite{ALPtriboson} placed bounds on the couplings to massive gauge bosons focusing on triboson final states at the LHC. Furthermore, LEP provides a good environment to directly test the coupling $g_{a\gamma Z}$ for ALP masses less than $m_Z$, as shown in Fig.~\ref{fig:exclusion-gaxaz}. The first constraint was placed in Ref.~\cite{EFTandCollider} by exploiting the limit on the uncertainty of the total $Z$ width. Another constraint on $g_{a\gamma Z}$ for $m_a$ in the range from 10GeV up to the $Z$ mass, labelled ``$Z\rightarrow \gamma+had$", was obtained in Ref.~\cite{ALPbosons}. However, this strong constraint is based on an assumption $g_{agg} >> g_{a\gamma Z}$, which will be released for a tiny gluonic coupling $g_{agg}$. Additionally, non-resonant production of two bosons via the gluon-gluon fusion (``non-resonant ggF")~\cite{nonresggF} and non-resonant vector boson scattering (``non-resonant VBS")~\cite{nonresVBS} can provide important constraints on $g_{aZZ}$ and $g_{a\gamma Z}$. However, the ``non-resonant ggF" constraints can be lifted completely for $g_{agg} \rightarrow 0$ .

In conclusion, the future $Z$ factories can provide a very good environment to test the couplings $g_{aZZ}$ and $g_{a\gamma Z}$ by exploring four signal channels. Compared to the LHC, the future $Z$ factories are more sensitive to $g_{a\gamma Z}$ ($g_{aZZ}$) via the $Z\rightarrow e^+ e^- \mu^+\mu^-$ and $Z\rightarrow e^+ e^- bb$ ($Z\rightarrow \mu^+\mu^- \slashed E$ and $Z\rightarrow bb\slashed E$ ) channels for $m_a$ in the range from 5GeV to tens GeV. It is important to note that the four channels probe the ALP interactions with EW bosons directly and independently of the coupling to gluons as ``non-resonant VBS" limits.
 It is expected that the future $Z$ factories could discovery or exclude ALPs with $m_a$ in the range of 5GeV -- 70 GeV.

\section*{Acknowledgements}

This work was supported in part by the National Natural Science Foundation of China under Grants No. 11875157, No. 11875306 and No. 12147214. SY would like to thank Yi-Lei Tang for helpful discussions.

\end{document}